\documentclass[%
 reprint,
 amsmath,amssymb,
 aps,
]{revtex4-1}

\usepackage{graphicx}
\usepackage{dcolumn}
\usepackage{bm}

\begin{document}

\preprint{APS/123-QED}

\title{FROM QUANTUM TO CLASSICAL WITHOUT $\hbar\rightarrow 0$}

\author{E. Gozzi}
\affiliation{Department of Physics ,
Theoretical Section, University of Trieste, Strada Costiera 11, 34151 Trieste, Italy.\\
Istituto Nazionale di Fisica Nucleare, Sezione di Trieste, Italy.}





\date{\today}

\begin{abstract}
In this paper we put forward some simple rules which can be used in order to pass from the {\it quantum} Moyal evolution operator to the {\it classical} one of Liouville without taking the limit of $\hbar\rightarrow 0$. These rules involve the averaging over some auxiliary variables.
\begin{description}
\item[PACS numbers] 03.65.-W, 5.20.-y
\end{description}
\end{abstract}

\maketitle


\section{\label{sec:level1}Introduction}
The  interplay between quantum (QM) and classical mechanics (CM) has been a topic of research since almost a century. It has been investigated both in the direction of going from CM to QM (quantisation) and in the opposite direction from QM to CM (which we like to call "dequantisation"\cite{dequa}). This is not the "semiclassical" limit like the one we obtain via $\hbar\rightarrow 0$. This last limit in fact is rather singular and moreover does not capture entirely CM because of the presence of phases which survive in their semiclassical limit. Moreover it does not capture the fact that  the real degrees of freedom of CM are objects much bigger than the Planck cells . There is a sort of "{\it coarse graining}" in phase-space that has to be performed in order to really pass from QM to CM. This was first indicated in the nice book by Peres \cite {isr} and further pursued  in ref.\cite {vien} and maybe in other papers which have escaped our attention.

\par
In this paper of ours we want to implement "dequantisation"  in a manner different than the one we experimented via path-integrals  in ref.\cite{dequa} . This new method is operatorial and  brings to light some sort of  "coarse graining". Unfortunately this last procedure is on some auxiliary variables whose  physical interpretation we do not fully understand  at the moment.

The operatorial method we shall use  is the one of the Moyal-Weyl-Wigner formalism \cite{mowewi} once it is brought in operatorial form like it was done in ref.\cite{mago}.

\par The paper is organised as follows: in Section 2) we give a brief review of ref.\cite{mowewi} using the notation of ref.\cite{mago}. In Section 3) we will provide the set of rules which brings us directly from the Moyal {\it quantum}  evolution to the {\it classical } Liouville one. These are the "dequantisation rules" we talked above and  they do not involve the limit of $\hbar\rightarrow 0$. In the last section ( the no. 4) we will comment on the kind of "coarse graining" we had to perform with the rules given above.

\section{\label{sec:level1}Brief Review of the Moyal Formalism}

This is a formalism which makes use only of functions on phase-space instead of operators on an Hilbert space. Let us start with a  space ${\mathcal{M}}={\mathcal{M}}_{2N}$ which is the $2N$ dimensional phase-space of a classical system. We will indicate the local coordinates in ${\mathcal{M}}_{2N}$ with $\varphi^a=(q^1 ... q^N ; p^1 ... p^N)$ where $a=1 ... 2N$.  The Hamilton equations of motion can be written as \cite{mars}:

\begin{equation}
{\dot\varphi^a}=\omega^{ab}\frac{\partial H}{\partial \varphi^b}
\label{due-uno-p}
 \end{equation}

 \noindent where $H$ is the original Hamiltonian of the system and locally $\omega^{ab}$ is a 2N dimensional matrix of the form 
 \begin{equation}
\left(
\begin{matrix}
0&\mathbb{I}_N\\
-\mathbb{I}_N&0\\
\end{matrix}
\right).
\label{due-due-p}
\end{equation}

This matrix is called symplectic matrix and via its inverse $\omega_{ab}$, once it is defined globally,  it is possible to build the so called symplectic two form $\omega$ \cite{mars} defined as : 
 \begin{equation}
\omega=\frac{1}{2}\,\omega_{ab}\,d\varphi^a\wedge d\varphi^b.
\label{tre-uno-p}
\end{equation}

The Moyal formalism begins by associating a function $O(q,p)$ on phase space to a quantum operator  $\widehat O({\hat q},{\hat p})$ defined on an Hilbert space.This association is called the $"{\it symbol}"$ map and it is indicated as :
\begin{equation}
O(\varphi^a)=symb(\widehat O).
\label{tre-due-p}
\end{equation}

\noindent Its definition  is:

\begin{equation}
O\left(\varphi^a\right)\equiv\int \frac{d^{2N}\varphi_0^a}{(2\,\pi\,\hbar)^N}\, \exp\left[\frac{i}{\hbar}\varphi^a_0\,\omega_{ab}\,
\varphi^b\right]\cdot Tr\left[\widehat T\left(\varphi^a_0\right)\,\widehat O\right].  \label{tre-tre-p}
\end{equation}

\noindent where

\begin{align}
\widehat T\left(\varphi^a_0\right)&\equiv \exp\left[\frac{i}{\hbar} \widehat{\varphi}^a\,\omega_{ab}\,\varphi^b_0\right]\nonumber\\
&=\exp\left[\frac{i}{\hbar}\left(p_0\,\hat q-q_0\,\hat p\right)\right] 
\label{quattro-uno-p}
\end{align}

\noindent and $\varphi^a_0$ is a fixed point in phase-space. There are various symbols associated to different ordering of $\hat q, \hat p$ in the operators. The one we have defined above is called Weyl symbol.

This procedure applies also to that operator called the density matrix $\widehat{\rho}$ which, for a pure state $|\psi>$,  is  $\widehat{\rho}=\vert\psi><\psi\vert$ and its  symbol $\rho$  is  known as the Wigner function, so  $\rho=symb \,{\widehat\rho}$, (see ref.\cite{mowewi} for more details).

\par
The non-commutative character of  the {\it operators} of QM is inherited by the {\it functions} of the Moyal formalism via a non-commutative product called star-product "$\star$" which has the following main property:

\begin{equation}
symb({\widehat A}{\widehat B})=symb (A) \star symb (B).
\label{quattro-due-p}
\end{equation}

\noindent Its explicit definition is:

\begin{align}
\left(A\ast B\right) \left(\varphi \right) &=A\left(\varphi\right)\,\exp\left[\frac{i\,\hbar}{2}\,\overset{\leftarrow}{\partial_a}\,\omega^{ab}\,\overset{\rightarrow}{\partial_b}\right]\,B\left(\varphi\right)\\
&=A(\varphi) B(\varphi) +0(\hbar).
\label{cinque-uno-p}
\end{align}

This is somehow a {\it "quantum deformation"} of the usual commuting product among functions. We can also build the symbol of the commutators of two operators and this symbol is called Moyal bracket (mb):

\begin{equation}
\left\{A,B\right\}_{mb}\equiv \frac{1}{i\,\hbar}\left(A\ast B-B\ast A\right)=\textrm{\emph{symb}}\left(\frac{1}{i\,\hbar}\left[\hat A, \hat B\right]\right).
\end{equation}

\noindent Its explicit expression is: 

\begin{align*}
\left\{A,B\right\}_{mb}&=A\left(\varphi\right)\,\frac{2}{\hbar}\sin\left[\left(\frac{\hbar}{2}\right)\,\overset{\leftarrow}{\partial_a}\,\omega^{ab}\overset{\rightarrow}{\partial_b}\right] \,B\left(\varphi\right)\\
&=\left\{A,B\right\}_{pb} + O\left(\hbar^2\right).
\end{align*}

\noindent We can consider the Moyal brackets (mb) a quantum deformation of the Poisson Brackets (pb) which is  the first term on the RHS of the second line of the expression above.
\vskip 2mm
The Heisenberg equation for the evolution of the density matrix which is :
\[
i\,\hbar\partial_t\widehat{\rho}=-\left[{\widehat \rho},\hat H\right]
\]
via the \emph{symbol map}, goes into the following equation
\begin{equation}
i\,\hbar\partial_t \rho(\varphi,t)=-\left\{\rho,H\right\}_{mb}
 \label{cinque-due-p}
\end{equation}
where $\rho$ and $H$ are the symbols of respectively $\widehat{\rho}$ and $\hat{H}$. For pure states this is the equation of motion for the Wigner functions.  
In the classical limit $\hslash\rightarrow0$ the  equation (\ref{cinque-due-p}) becomes
\begin{equation}
\partial_t \rho(\varphi,t)=-\left\{\rho,H\right\}_{pb}  
\label{sei-uno-p}
\end{equation}
which can also be written in the following operatorial form 
\begin{equation}
i \partial_t \rho(\varphi,t)= \hat L\,\rho  
\label{sei-due-p}
\end{equation}
where  $\hat L$ is known as the classical  Liouville operator and has  the form:
\[
\hat L= i \frac{\partial H}{\partial q} \frac{\partial}{\partial p}- i \frac{\partial H}{\partial p} \frac{\partial}{\partial q}  = - i \omega^{ab} \frac{\partial H}{\partial \varphi^b} \frac{\partial}{\partial \varphi^a}.
\]

Analogously to what has been done in eqs. (\ref{sei-uno-p}) and (\ref{sei-due-p}), we can perform for eq. (\ref{cinque-due-p})
\begin{eqnarray}
i\hbar\partial_{t}\rho & = & -\left\{ \rho,H\right\} _{mb} \nonumber \\
 & \Downarrow \nonumber \\
i\partial_{t}\rho & = & \widehat{L}_{\hbar}\rho \label{sei-tre-p}
\end{eqnarray}
where
\begin{eqnarray}
\widehat{L}_{\hbar} & = & - i\sum_{n=0}^{\infty}\frac{2}{\hbar}\frac{\left(-1 \right)^n}{\left(2n+1\right)!}\left[\frac{\hbar}{2}\omega^{ab}\partial_{b}H\partial_{a}\right]^{2n+1} \label{sei-quattro-p}
\end{eqnarray}
and this object is often called {\it Quantum Liouville operator}. Expanding in $\hbar$ we get
\begin{eqnarray*}
\widehat{L}_{\hbar} & = & \widehat{L}+i\frac{1}{24}\hbar^{2}\left(\omega^{ab}\partial_{b}H\right)\left(\omega^{il}\partial_{l}H\right)\left(\omega^{jm}\partial_{m}H\right)\partial_{a}\partial_{i}\partial_{j}
+O\left( \hbar^3 \right) 
\end{eqnarray*}
and so:
\begin{equation*}
\lim_{\hbar \rightarrow 0} \widehat{L}_\hbar =\widehat{L}.
\end{equation*}

At the classical level we can build the evolution operator associated to the classical eq. (\ref{sei-due-p}):
\begin{equation}
\widehat{U}_c = e^{-i \widehat{L} t}  . 
\label{sette-uno-p}
\end{equation}

\noindent The path-integral  which can be derived \cite{dequa} from  this evolution operator is :
\begin{equation}
\widehat{U}_c \rightarrow \int {\cal{D}} \varphi^a {\cal{D}} {\lambda_a } 
e^{i \int \widetilde{\cal{L}}_B dt} 
\label{sette-due-p} 
\end{equation}

\noindent where $\varphi^{a}$  are  the usual phase-space variables and  the $\lambda_a$ are auxiliary variables defined in ref.\cite{dequa} while  the Lagrangian used in (\ref{sette-due-p}) is:

\begin{equation}
\widetilde{\cal{L}}_B = \lambda_a \dot{\varphi}^a -\lambda_a \omega^{ab} \frac{\partial H}{\partial \varphi^b}.
\label{sette-tre-p}
\end{equation}
 
Differently than in ref.\cite{dequa} we do not introduce here further auxiliary variables $(c^a, {\bar c}_a)$ of grassmannian character and we do that in  order to simplify things. The reason for the subindex "B" (for Bosonic) on the lagrangian above is exactly because no Grassmannian variables has been introduced.

\par
The reader could now ask if also the quantum evolution operator associated to eq.(\ref{sei-tre-p}), that is:

\begin{equation}
{\widehat U}_{q}= e^{-i {\widehat L}_{\hbar}t}
\label{sette-quattro-p}
\end{equation}

\noindent has a path-integral counterpart. The answer is yes and this was done by Marinov in ref. \cite{mago}. The result is:

\begin{equation}
\widehat{U}_q \rightarrow \int {\cal{D}} \varphi^a {\cal{D}} {\lambda_a } 
e^{i \int \widetilde{\cal{L}}^{\hbar} _Bdt} 
\label{otto-uno-p} 
\end{equation}

\noindent where $\widetilde{\cal{L}}^{\hbar}_B $ is :

\begin{equation*}
\widetilde{\cal{L}}^\hbar_B = \lambda_a \dot{\varphi}^a - \widetilde{{\cal H}}_{B}^{\hbar} \,.
\end{equation*}
\noindent with the "hamiltonian"  $ \widetilde{{\cal H}}_{B}^{\hbar}$  having  the form:

\begin{eqnarray}
\widetilde{{\cal H}}_{B}^{\hbar} & = & \frac{1}{2\hbar}\left[H\left(\varphi^{a}-\hbar\omega^{ab}\lambda_{b}\right)-H\left(\varphi^{a}+\hbar\omega^{ab}\lambda_{b}\right)\right]  \label{otto-due-p}.
\end{eqnarray}

\noindent The two $H$ appearing on the RHS of the equation above are the same Hamiltonian function of the original eq. (\ref{due-uno-p}) but with arguments  different than the simple $\varphi$.

To compare the expression (\ref{otto-due-p}) with the one of Marinov in ref. \cite{mago} one should keep in mind that Marinov used a variable $\xi^a$
which is related to ours in the following manner : $\xi^{a}=\omega^{ab}\lambda_b$.

In  the classical path-integral formalism of eq. ( \ref{sette-due-p}) we work in a sort of extended phase-space $(\varphi^a, \lambda_a)$ and  the Hamiltonian associated  to  eq.(\ref{sette-tre-p}) is given by:

\begin{equation}
\widetilde{{\cal H}}_{B}=\lambda_a\omega^{ab}\frac{ \partial H}{\partial \varphi^b}
\label{nove-uno-p}.
\end{equation}

\par
From the path integral expression (\ref{sette-due-p}) for classical physics  and the associated Lagrangian in
(\ref{sette-tre-p}), one can get the following commutators between the operators associated to $\varphi^a$ and $\lambda_a$:

\begin{equation}
[{\widehat\varphi}^a, {\widehat \lambda}_b]=i\delta^a_b.
\end{equation}

\noindent This implies that ${\widehat\lambda}_b=-i\frac{\partial}{\partial\phi^b}$ and the $\widetilde{{\cal H}}_{B}$ of eq.(\ref{nove-uno-p}) becomes the Liouville operator  of eq.(\ref{sei-due-p}): 
\begin{equation}
\widetilde{{\cal H}}_{B}\rightarrow -i\omega^{ab}\frac{\partial H}{\partial\varphi^b}\frac{\partial}{\partial\varphi^a}\rightarrow {\hat L}
\label{nove-due-p}
\end{equation}

\section{\label{sec:level1}From Quantum To Classical without $\hbar\rightarrow 0$}

Our goal in this section is to find a set of rules which can bring us from    
the quantum $\widetilde{{\cal H}}_{B}^{\hbar}$  of (\ref{otto-due-p}) to the classical one $\widetilde{{\cal H}}_{B}$ of (\ref{nove-uno-p}) without taking the limit of $\hbar\rightarrow 0$.
\par
\vskip 3mm
\noindent {\bf 1)}\break 
The first rule is to multiply the $\omega^{ab}$ present in (\ref{otto-due-p})
by two grassmannian variables $\theta, \bar\theta$, and insert this new expression  into $\widetilde{{\cal H}}_{B}^{\hbar}$ :

\begin{equation}
\omega^{ab}\longrightarrow  \theta{\bar\theta}\omega^{ab}
\label{dieci-uno-p}
\end{equation}
\begin{equation}
\widetilde{{\cal H}}_{B}^{\hbar}(\varphi^a, \hbar\omega^{ab}\lambda_b))\longrightarrow \widetilde{{\cal H}}_{B}^{\hbar}(\varphi^a, \hbar\theta{\bar\theta}\omega^{ab}\lambda_b)
\label{dieci-due-p}
\end{equation}

\par
\vskip 3mm
\noindent {\bf 2)}\break
The second rule is to integrate  over  $\theta$ and $\bar\theta$
the $\widetilde{{\cal H}}_{B}^{\hbar}(\varphi^a,\hbar \theta{\bar\theta}\omega^{ab}\lambda_b)$
present on the RHS of eq.(\ref {dieci-due-p} ). The rules for integration over grassmannian variables are the following ones (see ref.\cite{bry}):
\begin{equation}
\int  \theta  d\theta= 1
\label{referee1}
\end{equation}
\begin{equation}
\int   d\theta= 0
\label{referee2}
\end{equation}

and the same ones by replacing everywhere above the $\theta$ with the variable $\bar\theta$.

\noindent We shall  prove that the result of the integration of point {\bf 2}) above is the following:

\begin{equation}
\int \widetilde{{\cal H}}_{B}^{\hbar} d{\bar\theta}d\theta=\widetilde{{\cal H}}_{B}
\label{dieci-tre-p}
\end{equation}

\noindent This implies that the set of two rules {\bf 1)} and {\bf 2)} lead us  from the quantum to the classical without $\hbar\rightarrow 0$.
\par
Let us now prove eq.(\ref{dieci-tre-p}). If you remember eq.(\ref {otto-due-p})
the $\widetilde{{\cal H}}_{B}^{\hbar}$ is made of two pieces:

\begin{eqnarray}
\widetilde{{\cal H}}_{B}^{\hbar} & = & \frac{1}{2\hbar}\left[H\left(\varphi^{a}-\hbar\omega^{ab}\lambda_{b}\right)-H\left(\varphi^{a}+\hbar\omega^{ab}\lambda_{b}\right)\right]  \label{undici-uno-p}
\end{eqnarray}

\noindent Let us apply the rule (\ref{dieci-uno-p}) to each of the two pieces on the RHS of the equation above. The second piece  is $H(\varphi^a+\hbar\theta{\bar\theta}\omega^{ab}\lambda_b)$ and if we expand it in $\theta$ and ${\bar\theta}$ we get 

\begin{equation}
H(\varphi^a+\hbar\theta{\bar\theta}\omega^{ab}\lambda_b)=F+\theta G+{\bar \theta}
L+\theta{\bar\theta}M
\label{undici-due-p}
\end{equation}

\noindent where $F,G,L,M$ are functions of $(\varphi,\hbar\omega^{ab}\lambda_b)$
which we will determine in what follows.

If we put $\theta$ and $\bar\theta$  to zero on both sides of eq.(\ref{undici-due-p})
we get:

\begin{equation}
H(\varphi)=F
\label{undici-tre-p}
\end{equation}

\noindent So $F$ is nothing else that the original Hamiltonian (\ref{due-uno-p}) of the system.  Next let us do the derivative with respect to $\theta$ on both side of eq.(\ref{undici-due-p}) and then put $\theta$ and $\bar\theta$ to zero. This is a simple calculation and leads to :

\begin{equation}
G(\varphi,\hbar\omega^{ab}\lambda_b)=0
\label{undici-quattro-p}
\end{equation}

In the same way, by doing the derivative with respect to $\bar\theta$ of both sides of eq.(\ref{undici-due-p})  and putting $\theta$ and $\bar\theta$ to zero we easily get:

\begin{equation}
L(\varphi,\hbar\omega^{ab}\lambda_b)=0
\label{undici-cinque-p}
\end{equation}

Finally let us do, on eq. (\ref{undici-due-p}) the second derivative , first with respect to $\bar \theta$ , next with respect to $\theta$, and put them to zero at the end. Simple calculations give the result:

\begin{equation}
M(\varphi,\hbar\omega^{ab}\lambda_b)=\hbar	\frac{\partial H(\varphi)}{\partial\varphi^a}\omega^{ab}\lambda_b
\label{undici-sei-p}
\end{equation}

If we now put together the equations: (\ref{undici-tre-p}), (\ref{undici-quattro-p}), (\ref{undici-cinque-p}) and (\ref{undici-sei-p}) we get :

\begin{equation}
H(\varphi^a+\hbar \theta{\bar\theta}\omega^{ab}\lambda_b)=H(\varphi)+\theta{\bar\theta}\hbar\frac{\partial H(\varphi)}{\partial\varphi^a}\omega^{ab}\lambda_b
\label{dodici-uno-p}
\end{equation}
If we do the same expansion in $\theta$ and $\bar\theta$ for the first Hamiltonian that we have on the RHS of eq. (\ref{undici-uno-p}) we get
\begin{equation}
H(\varphi^a-\hbar \theta{\bar\theta}\omega^{ab}\lambda_b)=H(\varphi)-\theta{\bar\theta}\hbar\frac{\partial H(\varphi)}{\partial\varphi^a}\omega^{ab}\lambda_b
\label{dodici-due-p}
\end{equation}

Inserting all  this in the Hamiltonian which appear on the RHS of eq.(\ref{dieci-due-p})  we obtain:

\begin{equation}
\widetilde{{\cal H}}_{B}^{\hbar}(\varphi,+\hbar \theta{\bar\theta}\omega^{ab}\lambda_b))=-\theta{\bar\theta}\frac{\partial H(\varphi)}{\partial\varphi^a}\omega^{ab}\lambda_b
\label{dodici-tre-p}
\end{equation}

Let us notice that the $\hbar$ has {\it miraculously} disappeared. Finally ,  using the relation (\ref{dodici-tre-p}), let us perform the second step contained in rule {\bf 2)} of eq.(\ref{dieci-tre-p}): 

\begin{align}
\int \widetilde{{\cal H}}_{B}^{\hbar}(\varphi^a,\hbar \theta{\bar\theta}\omega^{ab}\lambda_b)d{\bar\theta}d\theta&= \int -\theta{\bar\theta}\frac{\partial H(\varphi)}{\partial\varphi^a}\omega^{ab}\lambda_b d{\bar\theta}d\theta\\
&=\lambda_a\omega^{ab}\frac{\partial H}{\partial \varphi^b}=\widetilde{{\cal H}}_{B}
\end{align}

This is exactly the relation  (\ref{dieci-tre-p}) that we wanted to prove. 

So we can conclude that we have gone from quantum to classical without ever doing $\hbar\rightarrow 0$.

\section{\label{sec:level1}conclusion}
In this conclusion we will try to understand in a more physical way  what we have done. It will be  a primitive attempt anyhow.

With the step (\ref{dieci-uno-p}) we basically got a different manner to write the symplectic two-form which is usually written as \cite{mars}:

\begin{equation}
\omega= \omega_{ab}d{\varphi}^{a}\wedge d{\varphi}^{b}
\label{tredici-uno-p}
\end{equation}

\noindent where "$\wedge$" is the so called wedge product among forms and it is anticommuting \cite{mars}.

Via the substitution (\ref{dieci-uno-p}) we can now eliminate the symbol of wedge product and have (\ref{tredici-uno-p}) written as :

\begin{equation}
\omega= \omega_{ab}(d{\varphi}^{a}\theta) (d{\varphi}^{b}{\bar\theta} )
\label{tredici-due -p}
\end{equation}

Here we should add the rule that when we exchange $d\varphi^{a}$ with
$d\varphi^b$ we have to exchange also the $\theta$ with the $\bar\theta$.
This we indicate with the fact that we put each $d\varphi$ in a bracket with its own $\theta$ or $\bar\theta$ as if it were a single object. So $d\varphi^a$ becomes , in our language, a differential form only if combined with either $\theta$ or $\bar\theta$. Somehow it is as if the grassmannian variables $\theta$ and $\bar\theta$ were some "basic units" for the anticommuting character of forms. Step {\bf 2)} of our dequantisation procedure is given by the rule in eq.(\ref{dieci-tre-p}) which consist of integrating over $\theta,{\bar\theta}$. This may be the "coarse graining" we mentioned at the beginning. We know that $\omega^N$ is the infinitesimal volume of phase-space and integrating over it would mean integrating over all of phase-space. This is not what the author of ref.\cite{isr} wanted to do. He somehow wanted to smear the phase-space contained in Planck-cells and this is not an easy job. May it be that our integration over $\theta$ and $\bar\theta$ does exactly that ?. We do not have an answer to this question and more work is needed. Somehow one should investigate  the {\it physical} meaning of the integration over grassmannian variables which, up to know, has only been a mere mathematical procedure \cite{bry}. A thing to start from is that $d{\theta}d{\bar\theta}$ has the same dimension  as $\hbar$ \cite{dequa} and plays a role  also in ref.\cite{dequa} where we experimented with a different  dequantisation procedure.
Work is in progress on this issue.

 \begin{acknowledgments}
This work has been supported  by INFN (GeoSymQft , Naples and gruppo IV , Trieste ). \end{acknowledgments}

\bibliography{apssamp}
  
\end{document}